\documentclass[12]{article}
\usepackage[utf8]{inputenc}
\usepackage{amsmath}
\usepackage{amsfonts}
\usepackage[margin=1in]{geometry}
\usepackage[ruled,vlined]{algorithm2e}
\usepackage{graphicx}
\usepackage{cite}
\usepackage{authblk}
\usepackage{setspace}

\title{Data Driven Reaction Mechanism Estimation via Transient Kinetics and Machine Learning}

\author[1]{M. Ross Kunz}
\author[2]{Adam Yonge}
\author[1]{Zongtang Fang}
\author[1]{Rakesh Batchu}
\author[2]{Andrew J. Medford}
\author[3]{Denis Constales}
\author[4]{Gregory Yablonsky}
\author[1]{Rebecca Fushimi \footnote{Corresponding author: Rebecca Fushimi, rebecca.fushimi@inl.gov}}
\affil[1]{Energy $\&$ Environment Science and Technology, Idaho National Laboratory}
\affil[2]{School of Chemical $\&$ Biomolecular Engineering, Georgia Institute of Technology}
\affil[3]{Department of Electronics $\&$ Information Systems ELIS, Ghent University}
\affil[4]{Department of Energy, Environmental $\&$ Chemical Engineering, Washington University in St. Louis}

\date{\today}

\begin{document}

\maketitle

\doublespacing

\begin{abstract}
	Understanding the set of elementary steps and kinetics in each reaction is extremely valuable to make informed decisions about creating the next generation of catalytic materials. 
	With structural and mechanistic complexities of industrial catalysts, it is critical to obtain kinetic information through experimental methods. 
	As such, this work details a methodology based on the combination of transient rate/concentration dependencies and machine learning to measure the number of active sites, the individual rate constants, and gain insight into the mechanism under a complex set of elementary steps. 
	This new methodology was applied to simulated transient responses to verify its ability to obtain correct estimates of the micro-kinetic coefficients. 
	Furthermore, data from an experimental CO oxidation on a platinum catalyst was analyzed to reveal that Langmuir-Hinshelwood mechanism drives the reaction. 
	As oxygen accumulated on the catalyst, a transition in the apparent kinetics was clearly defined in the machine learning analysis due to the large amount of kinetic information available from transient reaction techniques. 
	This methodology is proposed as a new data driven approach to characterize how materials control complex reaction mechanisms relying exclusively on experimental data.
\end{abstract}

\newpage
\section{Introduction}

Intrinsic understanding of the reaction mechanism is highly sought after as it provides insight into the development and optimization of new catalysts. 
More specifically, the goal of extracting a set of governing reaction rules enables the design of new process schemes, the investigation of different operating regimes, and the ability to cope with deviations from steady-state operation \cite{matera2019Progress, medford2018extracting}.    
Widely used tools for understanding reaction mechanism are based on atomistic calculations, minimization of potential energy along the reaction coordinate, using density functional theory (DFT) followed by  Kinetic Monte Carlo (KMC) or Molecular Dynamics (MD) simulation \cite{steinfeld1989chemical, medford2015catmap, andersson2006toward, norskov2011density, norskov2002universality, reuter2011first}. 
In general, the assumptions regarding reaction intermediates required for each simulation are determined through stoichiometry, literature, and chemical intuition. 
Moving from the atomistic to reactor scale (a “bottom-up” methodology) increases the complexity of the simulation when estimating the performance of a particular catalyst with different coverages of adsorbed species. 
Arriving at experimentally verifiable models can be challenging due to the complexity of the catalyst structure, reaction mechanism, adspecies interactions, limitations in accuracy of atomistic simulations, and elementary reactions that occur on varying time scales \cite{ulissi2017address, medford2014assessing}. 

On the other hand, steady-state kinetic characterization methods are commonly used to derive mechanistic understanding from experimental rate and concentration data collected on the complex catalytic material. 
These methods typically provide apparent kinetic properties, limited to the resolution of the reactor device, which can be supported by multiple viable atomistic simulation results \cite{ambast2020passive, rangarajan2017sequential}. 
For example, the Continuously Stirred Tank Reactor (CSTR) and plug flow reactor (PFR) are widely used to measure the global behavior of the kinetics over a set of prepared catalysts and while steady-state experiments typically generate sparse sets of kinetic data that make fitting complex models less meaningful, transient experiments in these devices has been exploited to generate rich data sets \cite{miro1986continuous, aroh2018efficient, taylor2020rapid, waldron2020model, fath2020efficient, wojciechowski2003experimental}.   
Additionally, transient behavior may be measured within a fixed bed reactor using techniques such as Steady State Isotopic Transient Kinetic Analysis (SSITKA), Modulation Excitation Spectroscopy (MES) and Temporal Analysis of Products (TAP) \cite{ledesma2014recent, muller2017applications,constales2017precise, yablonsky2016rate, morgan2017forty}. 
In a steady state experiment the relationship between the gas and surface concentration is fixed. 
In a dynamic experiment, the two are decoupled and their dependencies reveal greater understanding for how the steps of the reaction mechanism work together.
The utility of transient data has been increasingly emphasized as the information can be collected over a variety of changing conditions such as concentration, pressure, and temperature over a function of time rather than a mean value. 
The large volume of dynamic information allows for detailed analysis of the catalytic behavior that supports more robust modeling and simulation. 
However, with an increase in the amount of chemical information provided by the transient experiment, there is also a burden of the mathematical techniques required for analysis to properly describe the interactions among chemical species.

The objective of this paper is to develop phenomenological reaction networks through quantitative transient kinetic characterization using a combination of well-defined reactor physics theory and data-driven machine learning analysis.  
Advances in machine learning and artificial intelligence driven by increased computational efficiency can help with complex data analysis. 
The application of machine learning to modeling and simulation from a bottom-up perspective has already proven useful  in a variety of tasks including prediction of material properties \cite{boes2016neural, boes2017neural, ulissi2017machine, kitchin2018machine}, estimating material structure information \cite{ulissi2016automated, back2019convolutional, palizhati2019toward} and enhancing the performance of simulation methods where complete analysis of the material is computationally intensive \cite{ulissi2017address, gomberg2017extracting, lei2019design, wang2018density, gao2018modeling, back2019toward, lym2019lattice, ludwig2007ab}. 
Less common is the application of machine learning where the objective is to drive fundamental understanding of the material from a top-down approach based on experimental data \cite{medford2018extracting, li2019designing, tian2019deriving, tian2020leveraging, tian2019predicting, tian2019estimating}. 

The top-down approach is typically more difficult in practice as the kinetic properties of the material must be estimated and the data sample space is typically low due the time investment in running the experiment, e.g. waiting to reach steady state, often leading to strategies that fuse information from computational techniques with experimental data. 
However, even in the presence of limited experimental data, the bioinformatics domain has demonstrated success from a top-down approach \cite{baxevanis2020bioinformatics, baldi2001bioinformatics}. 
For example, bioinformatics has been an early adopter for the analysis of high dimensional genomics and cancer information via machine learning for improvement in predictions in a low sample space \cite{breheny2011coordinate, zou2005regularization}. 
Given this example, our paper describes similar methodology for the application of catalysis informatics. 
We propose to combine the large volume of transient kinetic data with machine learning to find a new description of the underlying gas and surface interactions, i.e. the reaction mechanism. 

We begin by describing the transient Temporal Analysis of Products (TAP) reactor and how this instrument may be used to obtain time-dependent measurements of the reaction rate, gas concentration and surface uptake.  
The same strategy could be adopted by other transient techniques that provide similar time dependent data. 
The transient data is utilized in the proposed methodology, not as a prediction problem in machine learning, but as a high-dimensional problem where the objective is to create an interpretable and simplified representation of the reaction network. 
Through parsimonious feature selection with non-convex optimization and examination of the covariance structure, we provide quantitative information about a reaction mechanism. 
This methodology does not require any prior assumptions of the mechanism, making it applicable on a wide range of gas/surface reactions. 
This methodology was validated against simulated data with a known set of simple reactions and with respect to the Eley-Rideal and Langmuir Hinshelwood mechanisms. 
These simple mechanisms are useful in understanding how machine learning methods may be used to infer kinetics and the overall reaction mechanism . 
Furthermore, the developed methodology was applied to a series of TAP pulse responses within an experimental CO oxidation data set. 
This experiment provided a unique observation of the impact of evolving catalyst surface concentration on the reaction mechanism.

\section{Methodology}\label{sec:methodology}
Transient  experiments offer the ability to distinguish relevant kinetic information with respect to the set of elementary steps that support a global reaction. 
As such, many have used chemical intuition of the reaction pathway to form a set of elementary steps that can then be used to validate the behavior of transient experimental data \cite{reece2017kinetic, schuurman2007assessment}.  
Due to the pure volume of data and mathematical ties to the physics of the experiment, transient kinetic data is a prime candidate for the application of machine learning to discern the interplay of chemical species without \textit{a priori} kinetic assumptions, i.e. in a mechanistically model free manner \cite{medford2018extracting}. 
This section focuses on transient kinetic data obtained by the Temporal Analysis of Products (TAP) reactor where it will be shown how machine learning may be naturally augmented with the physics of this unique reactor to provide insight into the mechanism.

The Temporal Analysis of Products (TAP) reactor system is a transient method that can bridge the gap between well defined systems and complex structural or compositional industrial catalysts \cite{gleaves1988temporal, shekhtman1999thin}. 
The TAP reactor is an exclusively diffusion-driven reactor that offers the ability to measure the gas concentration over time with respect to changes in  the surface concentration of adsorbed species. 
Experiments yield an outlet flux within a fine time resolution, typically milliseconds, for each observable reactant and product gas phase response that is directly related to the conversion, yield and selectivity. 
Adding subsequent pulses allows the user to alter the catalytic state in a controlled manner, and to monitor how the kinetic properties shift throughout the experiment. 
As such, moment-based techniques that describe the apparent kinetics may be applied to each outlet flux response and can be interpreted equivalently to many steady state reactions \cite{zheng2008needle, assmann2003ruthenium, assmann2008heterogeneous, yablonskii1998moment}. 
While simple and straight-forward, using moment-based techniques or summary statistics negates the appeal of using TAP data as it discards the transient behavior where greater detail of the gas/surface interactions can be found \cite{wang2019rate}. 
For example, only using the first two moments in describing the transient response may not sufficiently describe the skewness, kurtosis, or even reversible adsorption of the response \cite{gleaves1997tap}.
The mathematics required to appropriately model the physics behind the reactor transport are described by Fick’s second law of diffusion. 
As the complexity of the reaction mechanism increases, so too does the system of coupled partial differential equations (PDEs) that must be optimized when fitting the transient response, though these PDEs can be simplified with carefully chosen reactor and experimental conditions. 
For  example, TAP experiments performed in the low-pressure regime enables well-defined Knudsen transport that is easily separated from the reaction, efficient mixing that ensures concentration uniformity in the bed of the catalyst and isothermal reaction conditions. 
This simplification eliminates advective or gas phase complications \cite{morgan2017forty, constales2016advanced}.

Taking advantage of a thin-zone catalyst configuration and Knudsen regime transport, the Y-Procedure is able to transform the TAP outlet response in the Laplace domain to directly measure the reaction rate and gas concentration as a function of time without any assumptions of the reaction mechanism \cite{yablonsky2007procedure, redekop2011procedure, kunz2018pulse}. 
Fundamentally, the Y-Procedure is a Fourier transform applied to the measured flux at the reactor outlet and the initial pulse at the inlet. 
Since the transport of the species through the inert region is well defined and only changes based on temperature and fractional voidage, the same formulation of the Fourier transform can be applied to trace the concentration and flux through the inert zone to the catalytic region. 
Alternatively, the same kinetic information can also be obtained in a statistical distribution fashion using the G-Procedure \cite{kunz2020probability}. 
Post-processing of the Y or G-Procedure also allows the user to determine the dynamic surface accumulation, or “uptake”, of each of the gas species. 
The uptake is defined as the integral of each reaction rate with respect to time scaled by stoichiometry, which yields the amount of each element that is contained in all adsorbed species \cite{redekop2014elucidating}. 
More specifically, the uptake/release, $U$, is defined as:
\begin{equation}
	U = \int^t_0 \sum_i v_i r_i^+ - \sum v_i r_i^- dt
\end{equation}
where $v$ is the stoichiometric coefficient, and $r$ is the reaction rate. 
When the reaction consists of a single elementary step, the uptake is equivalent to the surface concentration. 
However, when multiple elementary steps or types of active sites exist, the uptake represents dependent interactions between multiple gas and surface species.  
The time-dependent concentration, rate, and uptake are particularly useful because measurement of the terms requires no assumptions about the underlying mechanism.

\subsection{Rate expression as a linear model}
The transient chemical measurements are connected to the reaction mechanism by examining individual reaction expressions. 
More specifically, the transient rate, concentration and surface uptake may be combined to gain insight into ``reactivities'' that are related to the kinetic coefficients in a form described as the Rate-Reactivity model (RRM) \cite{yablonsky2016rate}. 
This differs from the common rate expression form where the rate of an elementary step is given as the interaction of the reacting species. 
For example, an irreversible rate expression of a gas interacting with the surface of the catalyst may use the following expressions to define the process:
\begin{align}
	\mbox{Elementary step} & \nonumber\\
	&A + * \to A^*\\
	\mbox{Rate expression} & \nonumber \\
	&r = k^+ C_A C_* \\
	\mbox{Rate expression realated to the number of surface sites}& \nonumber \\
	& r = \tilde{k}^+ C_A (N - C_{A^*}) \\
	\mbox{Rate-reactivity expression \cite{yablonsky2016rate}} & \nonumber \\
	& r = (k^+ N) C_A - k^+ C_A U_A \\
	\mbox{Linear model based on Rate-reactivity expression} & \nonumber \\
	& r = \beta_{(Nk^+)} C_A + \beta_{(k^+)} C_A U_A 
\end{align}
where $A$ is the gas species, $*$ denotes the surface species, $C_* (mol/m)$ is the concentration  of open active sites, $r$ $(mol/s)$ is the rate, $C (mol/m)$ is the gas concentration, $k (m^2/mol/s)$ is the positive definite kinetic coefficient, $N$ is the total number of active sites, $U (mol/m)$ is the surface uptake, and $\beta$ are the reactivity coefficients. 
The gas/surface interaction coefficient $\beta_{(k^+)}$ describes the intrinsic kinetic adsorption coefficient and the total number of active sites available on the surface may be determined through the ratio of $\beta_{(Nk^+)}$ and $\beta_{(k^+)}$ \cite{redekop2011procedure}.

The linear model form is a generalization of the rate expression where the dependent variable is the rate of the reactant while the independent variables consist of the gas concentration and surface uptake information. 
Each of the corresponding $\beta$ coefficients are the kinetic reactivity coefficients. 
Any higher-order gas/surface interaction that significantly impacts the reaction mechanism will include the polynomial interaction term ($CU$) denoting the element-wise multiplication of the gas concentration and the surface uptake while still being linear with respect to the rate. 
More explicitly, the full rate expression for second-order surface reactions is given by  
\begin{equation}\label{eq:rate_linear_model}
	r=\beta_C C + \beta_U U + \beta_{CU} CU + \beta_{CU^2} CU^2 + \beta_{U^2}U^2
\end{equation}
where each potential gas/surface interaction is displayed as one equation. 
Subsets of this linear model are described as traditional elementary steps in Table~\ref{tab:step_to_linear}. 
Each additional term in Equation~\ref{eq:rate_linear_model} grants further understanding about the kinetic coefficients. 
For example, $\beta_U$ is equivalent to the kinetic coefficient of desorption, $\beta_{(CU^2)}$ is the kinetic coefficient for the second order surface interaction, and $\beta_{(U^2)}$ is the kinetic coefficient of two surface  species that contribute to a gas concentration. 

\begin{table}[h!]
	\centering
	\begin{tabular}{c || c c c|} 
		\hline
		& Elementary Step & Rate Expression & Linear Form  \\
		\hline\hline
		1 & $A + * \to A^*$ & $r = k^+ C_A C_*$ & $r_A = \beta_{(Nk^+)} C_A - \beta_{(k^+)} C_A U_A$ \\ 
		2 & $A + * \rightleftharpoons A^*$ & $r=k^+ C_A C_* - k^-C_{A^*}$ & $r_A = \beta_{(Nk^+)}C_A -\beta_{(k^+)} C_AU_{A^*} - \beta_{(k^-)} U_{A^*}$ \\
		3 & $A+2* \to 2A^*$ & $r = k^+ C_A C_*^2$ & $r_A = \beta_{(Nk^+)} C_A - \beta_{(2Nk^+)} C_A U_{A^*} + \beta_{(k^+)} C_A U_{A^*}^2$ \\
		4 & $A^* + B^* \to 2* + C$ & $r = k^+C_{A^*}C_{B^*}$ & $r_{A^*} = \beta_{(k^+)}U_{A^*}U_{B^*}$ \\
		\hline
	\end{tabular}
\caption{\label{tab:step_to_linear}Comparison of elementary steps that correspond to the traditional rate expression and linear model coefficients.}
\end{table}

With application of the linear form to the TAP reactor, the units of the kinetic terms are reported at the catalyst thin zone where the rate is given as $mol/s$, concentration as $mol/m$, the adsorption rate constant as $m^2/mol/s$ and the desorption rate constant as $m/s$.  
As such, this can be directly applied to the units of the linear model coefficients. For example, a reversible reaction will have the following units:
\begin{align}
	r_A (mol/s) &= k^+ (m^2/mol/s)C_A(mol/m) \nonumber \\
	&- k^+ (m^2/mol/s)C_A(mol/m)U_{A^*}(mol/m) \\
	&- k^-(m/s) U_{A^*} (mol/m) \nonumber
\end{align}
which result in the following linear term coefficient units:
\begin{align}
	&\beta_{(Nk^+)} = m/s \\
	&\beta_{(k^+)} \mbox{ and } \beta_{(k^-)} = mol/s
\end{align}
From the thin-zone configuration, the following scaling terms may be applied to experimental data to transform these values into more commonly used units: the rate divided by the catalyst surface or grams of catalyst ($mol/m^2/s$ or $mol/g/s$) and the concentration divided by cylinder cross sectional area ($mol/m3$).

\subsection{Rate expression variable selection} \label{sec:variable_selection}
The full linear model in Equation~\ref{eq:rate_linear_model} describes all the single potential gas/surface interactions for mechanisms with second-order surface reactions, which encompasses the vast majority of mechanisms in heterogeneous catalysis \cite{laidler2013reaction}. 
As complexity of the elementary step and mechanism increases, manually fitting to the rate/concentration dependence becomes infeasible and may be dictated by potential biases. 
As such, we turn to variable selection within a machine learning framework for non-greedy optimization and selection of the kinetic coefficients.

The problem then becomes solving for the coefficients in the presence of real experimental data and determining which of the coefficients will be exactly zero. 
Let $\mathbf{y}$ be the response or dependent variable with dimensions $\bf{y} \in \mathbb{R}^{n \times 1} $ where n is equal to the number of time points collected. 
Then there exists a matrix  $\bf{X} \in \mathbb{R}^{n \times p} $, called the predictor matrix, where the total number of columns $p$ consists of the gas concentration, surface, and gas/surface interactions. 
From Equation~\ref{eq:rate_linear_model}, the problem in matrix form is defined as:
\begin{equation}
	\bf{y} = \bf{X} \boldsymbol{\beta} +\epsilon
\end{equation}
where $\epsilon$ is assumed to be Gaussian noise after correction of any cyclic noise due to electronic interference \cite{roelant2007noise}. 
The coefficients may be solved through optimization of $\boldsymbol{\beta}$, i.e., minimizing the sum of squared differences between the reaction rate and the predictor matrix:
\begin{equation}\label{eq:ols}
	\min_{\boldsymbol{\beta}} ||\bf{y} - \bf{X} \boldsymbol{\beta}||^2_2 
\end{equation}
where $||\cdots||^2_2$ denotes the 2-norm or Euclidean distance and it is assumed that each variable is mean centered so that the intercept is zero. 
Equation~\ref{eq:ols} is the general form of Ordinary Least Squares (OLS) model and is solved by setting derivative with respect to $\boldsymbol{\beta}$ to zero. There are subtle differences in the problem setup for the rate-reactivity model compared to the traditional least squares model. 
The objective is not to build a model to facilitate the prediction of the reaction rate, since that information will always be measured. 
Rather, the emphasis is placed on the ability of the regression model to correctly estimate the coefficient weights, or lack thereof, when a gas/surface concentration does not have an impact on the reaction mechanism. 
This requires that the regression model be able to provide consistent estimation of the coefficients and to perform variable selection of the concentrations in an autonomous fashion.

When the mechanism is hypothesized, the rate expression, and hence the predictor matrix, may be tailored to include select gas/surface interactions for estimation of the intrinsic kinetic coefficients. 
Biased models from hypothesized mechanisms, or statistical methods  that improperly select terms, will result in poor prediction accuracy of the reaction rate as well as an incomplete or incorrect description of the reaction mechanism. 
As such, model pruning through step-wise selection of the $p$-value or proper variable selection must be used.
Solving Equation~\ref{eq:ols} using standard statistical software results in $p$-values for each variable used in the model determining the relevance of the variable with respect to the rate. 
In a step wise fashion, variables can be added or removed based on the significance of the obtained $p$-value.
However, this method of variable selection is ‘greedy’ and may produce misleading statistically significant results based on the sample size. 
As an alternative, Equation~\ref{eq:ols} may be modified by adding a penalty to enforce restrictions on the size of the coefficients:
\begin{equation}\label{eq:ols_penalty}
	\min_{\boldsymbol{\beta}} ||\bf{y} - \bf{X} \boldsymbol{\beta}||^2_2 + P(\boldsymbol{\beta}, \lambda)
\end{equation}
where $\lambda$ is the tuning parameter. 
A common choice of the penalty for variable selection is to penalize the regression coefficients by the 1-norm, i.e. set $P(\boldsymbol{\beta},\lambda)_{LASSO}=\lambda|\boldsymbol{\beta}|_1$, known as the Least Absolute Shrinkage and Selection Operator (LASSO) \cite{tibshirani1996regression}. 
The 1-norm penalty in LASSO enforces shrinkage and exact zero coefficients for the values of $\beta$. 
When the value of $\lambda$ is zero, the OLS solution is obtained, and any increase in $\lambda$ will shrink the regression coefficients by $\lambda$. 
When the absolute value of the regression coefficient is less than the value of $\lambda$, the coefficient is set to zero. 
The value of $\lambda$ is typically tuned using $k$-fold cross validation which seeks to minimize the error of the model by using k random permutations of test and training subsets of the data \cite{hastie2009elements}. 
Ideally, the data would include a third prediction subset to verify that new predictions are robust, but this is not required as the focus in using a machine learning approach is inference on the regression coefficients that in turn describe the kinetic coefficients and not on the prediction.   
One drawback from the use of the 1-norm as a penalty is that the coefficients for the gas/surface terms may be significantly different in magnitude due to the units. 
Scaling each gas/surface variable by their respective standard deviation will result in the variables being of the same scale, but this would also alter the importance of variables representing the reaction.

An alternate penalty to the 1-norm is to use Smoothly Clipped Absolute Deviation (SCAD) where discontinuous shrinkage is enforced on the coefficients of the rate-reactivity model \cite{fan2001variable}. 
The penalty in Equation~\ref{eq:ols_penalty} may be replaced by the following function for SCAD:
\begin{equation}\label{eq:scad}
	P(\beta_i; \lambda, a)_{SCAD} = \begin{cases}
		\lambda | \beta_i|, & \mbox{ if } |\beta_i| \leq \lambda \\
		- \frac{1}{2(a - 1)} \left( |\beta_i| - 2 a \lambda |\beta_i| + \lambda^2 \right), & \mbox{ if } \lambda < |\beta_i| \leq a \lambda \\
		\frac{1}{2} (a +1) \lambda^2, & \mbox{ if } |\beta_i| > a \lambda
	\end{cases}
\end{equation}
where SCAD requires two different tuning parameters; $a$ and $\lambda$. 
The SCAD penalty is constructed in this way such that coefficients larger than $\lambda$ are not penalized while smaller values are still shrunk by $\lambda$ or set to zero. 
This allows the direct estimation of the rate reactivity coefficients as it can account for the discrepancy in the unit size in the gas and surface concentration variables. 
As a trade-off with the LASSO implementation, SCAD will require more computation time as it is no longer convex in optimization and requires two different penalty parameter terms. 
However, to relieve some of the computational burden, Fan et al. remark in the formulation of SCAD that $a$ can be fixed at a value of 3.7 through minimizing Bayes risk while $\lambda$ is tuned by $k$-fold cross validation \cite{fan2001variable}. 
In practical applications to TAP data, this difference between computation time is minimal. 
For example, when examining data with 5000 time points, SCAD runs within half a second compared to a tenth of a second when running LASSO at a collection time resolution of 0.001 seconds over 5 seconds on a 2.5GHz Intel core i7 processor.

\subsection{Elementary steps as a covariance structure}\label{sec:methods_cov}
As stated earlier, the rate-reactivity model uses data from the catalyst uptake or release detected in the gas outlet flux of a TAP pulse response experiment. 
This does have specific consequences when attempting to estimate the exact intrinsic kinetic coefficients with multiple active sites or elementary steps. 
Since the uptake is not equivalent as the catalyst surface coverage of a given species, the coefficients from the linear model are no longer the coefficients  associated with the traditional micro-kinetic model, but a combination that may be described as ``reactivities'' \cite{yablonsky2016rate}. 
This requires specific assumptions of the set of elementary steps to determine the exact mapping from the reactivities to the intrinsic kinetic coefficients.

Another approach to infer the mechanism without any \textit{a priori} information is to measure the relationship between each gas species by the joint variation, i.e. the covariance matrix. 
This again requires the assumption of being linear with respect to the rate, which is provided by the TAP experimental conditions, but also provides a complete view of the individual interactions between each gas and surface species. 
The problem in Equation~\ref{eq:rate_linear_model} changes from estimating the linear relationship between the rate, gas concentration, and surface concentration to finding the relationship of the rate and gas concentration dependencies of each gas species.  
More specifically, let each gas species be represented as the quotient of the reaction rate and the gas concentration ($Q_X = r_x / C_x$), then the covariance between gases $A$ and $B$ is given as:
\begin{equation}
	\mbox{cov}(A,B) = E\left[ \left( Q_A - \bar{Q}_A \right) \left( Q_B - \bar{Q}_B \right) \right]
\end{equation}
where $E$ is the expected value and $\bar{\cdot}$ denotes the mean \cite{casella2002statistical}. The Rate-Concentration Dependency Covariance structure (RCDC) describes the relationship between each surface uptake and hence may be leveraged to determine a specific reaction mechanism. 
As a concrete example, consider the Eley-Rideal mechanism where the first elementary step in a reaction mechanism is an irreversible process of gas A, i.e. $r_A = \beta_{(Nk^+_A)} C_A - \beta_{(k^+_A)} C_A U_A$, and the second elementary step, involves irreversible reaction of gas $B$ with adsorbed  $A$, i.e. $\beta_{(k^+_B)}C_B U_A$, then the covariance $Q_A$ and $Q_B$ describes the relationship between the dependent surface uptakes of $A$ and $B$. 

While the intrinsic kinetic coefficients may only be derived in very simple mechanisms via the linear model, the correlation between quotients can still be used to understand mechanistic features. 
When the covariance structure is normalized by the individual variances, this results in a correlation structure where the off diagonal values range from -1 to 1 depending on the intrinsic kinetic coefficients. 
As such, when the correlation between two different gas species is negative, there exists a parasitic interaction between the individual quotients and hence there exists a gas to surface relationship. 
This can be seen in the example of the Eley-Rideal mechanism, where the surface species from gas $A$ is being consumed by the concentration interaction of gas $B$, i.e., $corr(r_A/C_A, r_B / C_B) < 0$.  
Conversely, when the correlation coefficient is positive, there exists a mutualistic relationship where the individual gases may be both contributing to fill the available sites on the catalyst, i.e., $corr(r_A/C_A, r_B / C_B) > 0$.  
This can be extended to a mechanism as the correlation of reactants and products determines the elementary step relationships.  
For example, if $A$ and $B$ are both being consumed by the catalyst surface to form a product $C$ as in a Langmuir-Hinshelwood mechanism, then both correlation coefficients of $A$ and $B$ with respect to $C$ will be negative, i.e., $corr(r_A/C_A, r_C)$ and $corr(r_B / C_B), r_C) < 0$. 
This is in direct contrast to an Eley-Rideal mechanism where $B$ only interacts with the surface species of $A$ and hence $corr(r_A/C_A, r_C) < 0$ and $corr(r_B / C_B), r_C) > 0$.
As such, the sign of the correlation determines how the reactants and products interact via the catalyst surface.

As the total number of different rate-concentration dependencies increase based on the number of reactants and products in a mechanism, it can be assumed that some of the reaction components do not have a specific relation to one another. 
This can be thought of as an extension of the variable selection in Section~\ref{sec:variable_selection} within a reaction expression but applied to the sample covariance matrix to determine a sparse set of interaction between each rate-concentration dependency. 
These techniques are typically noted as sparse graph estimation where the objective is to provide a sparse representation of the inverse covariance matrix \cite{friedman2008sparse, fan2016overview, meinshausen2006high}. 
The combination of sparseness between rate-concentration dependencies and the mutualistic/parasitic relationship may be used to determine what species have interactions and what surface interactions are apparent.
This leads to a data driven method for deriving mechanistic information from complex, industrial catalytic materials that is distinct from \textit{a priori} assumptions of a conventional micro-kinetic model.

An algorithm for the calculation of the Rate-Concentration Dependency Correlation (RCDC) structure is given below. 
Each gas flux must be preprocessed by considering the mass spectrometer calibration factor for appropriate scaling of the number of molecules injected and the baseline of the flux is centered at zero \cite{kunz2018pulse}.  
Next, the Y or G-Procedure is applied to the corrected flux to obtain the time—dependent reaction rate, gas concentration, and uptake without any mechanistic assumptions. 
The rate/concentration dependency is calculated as the reaction rate and gas concentration ratio if the flux is a reactant. 
If the flux is a product, only the reaction rate is used as the rate/concentration dependency as the outlet flux is only formed from interaction with the catalyst. 
Finally, the RCDC for each combination of reactant and product is measured by taking the correlation of the rate/concentration dependencies.

\begin{algorithm}[H]
	\KwData{Let $F$ represent the gas flux, the mass spectrometer calibration coefficient as $\mu$, and the average baseline value of the flux as  $\bar{F}_b$}
	\ForEach{ flux $(i)$}{
		$ \hat{F}_i = \mu_i \left(F_i - \bar{F}_{b,i} \right) $ \;
		$r_i,C_i, F_i = Y(\hat{F}_i)$ \;
		\eIf{ $F_i$ is a reactant}{
			$RCD_i = r_i/C_i$\;
		}{
			$RCD_i = r_i$\;
		}
	}
	\ForEach{reactant $(i)$ and prodcut $(j)$}{
		$RCDC_{i,j} = corr(RCD_i, RCD_j)$
	}
	\caption{Rate-Concentration Dependency Correlation (RCDC)}
\end{algorithm}

\section{Data generation and experimental methods }

Simulated data using a thin zone TAP reactor was used to directly relate the linear model to the micro-kinetic coefficients. 
The simulated reactor configuration was normalized to unit values as much as possible to provide clear comparisons. 
The reactor length was set to a value of 1 $(m)$ where two inert beds of equal length encapsulate an infinitesimally small catalyst zone. 
The bed porosity of the inert material was set to 0.5 with diffusion coefficient of 1 $(m^2/s)$. 
A total of 1 $mol$ was injected into the reactor and the outlet flux was simulated out to 3 seconds with a time resolution of 0.001 second. 
Without any reaction present, the outlet flux under these reactor configurations results in the dimensionless Standard Diffusion Curve \cite{gleaves1997tap}. In the presence of reaction, the system of two ordinary differential equations modeling the deactivation,i.e., consumption of sites, of the catalyst was integrated using the adaptive Adams and BDF methods of the lsoda function (from ODEPACK) in the deSolve package of \textsf{R} \cite{soetaert2010deSolve,rCitation}.
The parameter estimation is performed using the modFit function of the \textsf{R} package FME \cite{soetaert2010FME}. 
To prepare the data, to handle the post processing, and to simulate TAP experiments, proprietary \textsf{R} code is used along with lsoda and modFit.

\subsection{Simulated data for validation of elementary step estimation}\label{sec:sim_data_elementary}
\begin{figure}[h!]
	\centering
	\includegraphics[width=.5\textwidth]{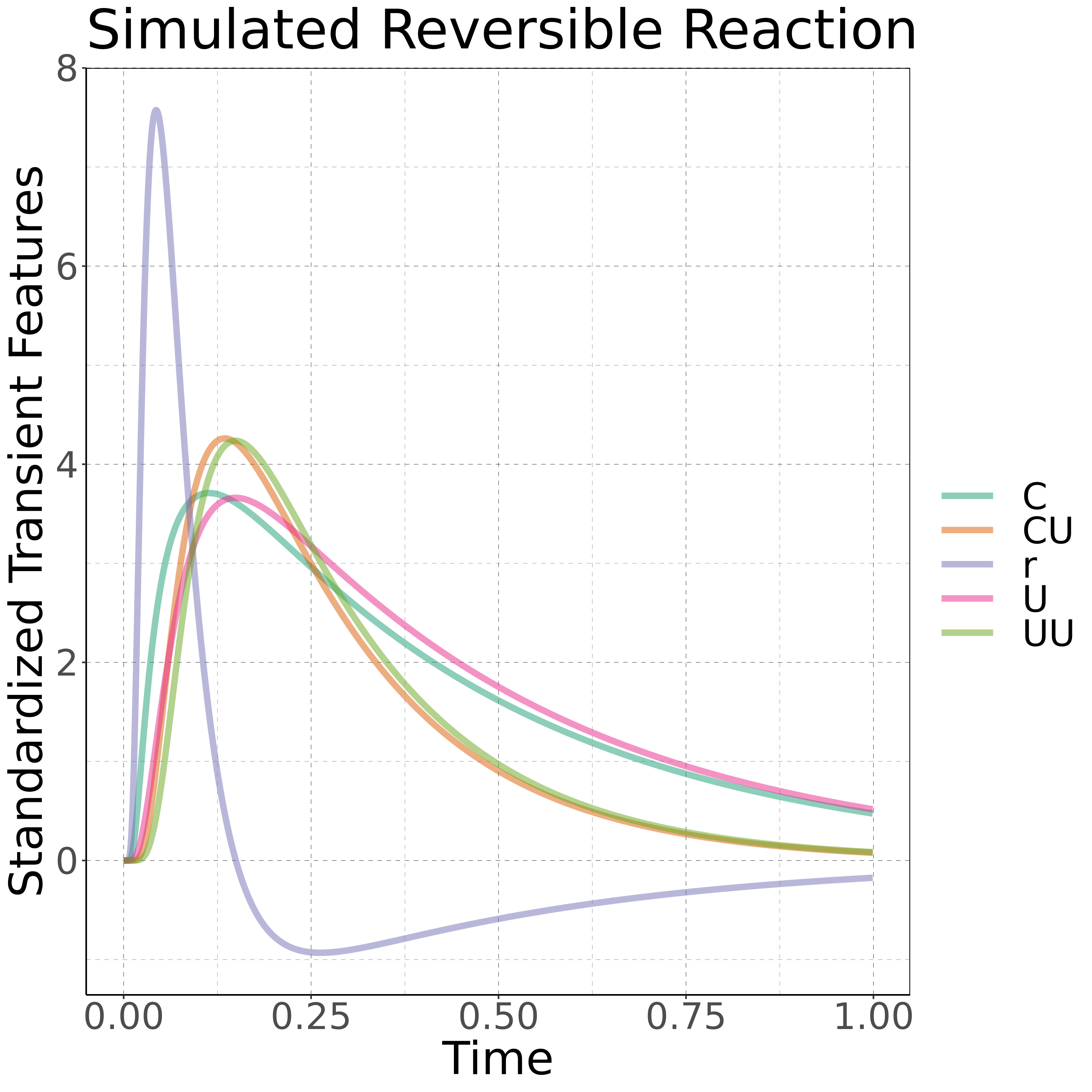}
	\caption{\label{fig:sim_flux}The simulated TAP transient features of $r$, rate, $C$, concentration and $U$, uptake for reversible reaction.} 
\end{figure}

To verify the accuracy of SCAD to correctly recover the kinetic coefficients, irreversible adsorption $A+*\to A^*$ and reversible adsorption $A+* \rightleftharpoons A^* $ were tested with different kinetic rate constants  and number of active sites. 
Recall from Table~\ref{tab:step_to_linear}, the linear form of both irreversible and reversible adsorption may be described as a single linear model, i.e. $r_A =k_{app}^+ + C_A U_{A^*} - k^- U_{A^*}$, where $k^-$ is zero when the reaction is irreversible. 
Three different cases of the linear model are tested: irreversible adsorption with an abundant number of active sites where $k_{app}^+$ dominates the reaction (row 1), irreversible adsorption where $k^-$ is zero (row 2 and 3), and reversible adsorption where all coefficients are present (row 4). 
Furthermore, an additional case was tested where the total number of active sites within irreversible adsorption differ.
This change in the number of active sites within irreversible adsorption are denoted as ($2a$) and ($2b$) within Table~\ref{tab:step_to_linear}.
All reactions did not include a second order surface uptake term.
Figure~\ref{fig:sim_flux} displays the simulated TAP kinetic features for reversible adsorption that were used within each regression model. 
The second order surface uptake feature is included in the figure, as well as in the regression model's potential variables, as spurious information to the reaction that should be automatically removed with variable selection. 
The coefficients within Table~\ref{tab:sim_coefficients} describe the complete set of different reactions tested. 
All simulations have an arbitrary fixed forward apparent rate constant ($k_{app}^+$) of 0.2 $m/s$  for ease of comparison between reactions. 

\begin{table}[h!]
	\centering
	\begin{tabular}{c || c c c c |} 
		\hline
		& Reaction & $\beta_{(Nk^+)}$ & $\beta_{(k^+)}$ & $\beta_{(k^-)}$ \\
		\hline\hline
		1 & Irreversible  (abundant sites) &0.2 & $-$ & 0\\ 
		2a & Irreversible (limited sites $N$=1) & 0.2 & $-0.2$ & 0 \\
		2b & Irreversible (limited sites $N$=2.5) & 0.2 & $-0.5$ & 0 \\
		3 & Reversible (limited sites $N$=1) & 0.2 & $-0.2$& $-40$ \\
		\hline
	\end{tabular}
	\caption{\label{tab:sim_coefficients}The set of simulated  irreversible and reversible reactions with different number of active sites.  $N$ is the normalized number of active sites.}
\end{table}

\subsection{Simulated data for comparison of the Eley-Rideal and Langmuir-Hinshelwood mechanisms}\label{sec:sim_data_mechanism}
The Eley-Rideal (ER) and Langmuir-Hinshelwood (LH) mechanisms will be compared with respect to carbon monoxide oxidation with the same reactor configuration above. 
When examining each mechanism as a set of irreversible steps, a grid of values for both the forward rate constant of oxygen and carbon monoxide was constructed ranging from 0.04 to 1.00 $m^2/mol\/s$ with an increment of 0.02 $m^2/mol\/s$ for each index. 
The grid of values was constructed to demonstrate the trends between the ER and LH mechanism when examining the correlation structure.
The forward rate constant of carbon dioxide was fixed at a value of 5.00 $m^2/mol\/s$ in the LH mechanism. Finally, to explore the impacts of reversibility on the correlation structure of the LH mechanism, the forward rate constant of oxygen was fixed at a value of 0.2 $m^2/mol\/s$ while the forward and reverse rate constant of carbon monoxide ranges in value from 0.04 to 1.00 $m/s$ with an increment of 0.02 $m/s$ for each index.

\subsection{Experimental data for application of the RCDC}\label{sec:experimental_data}
Experimental carbon monoxide oxidation on platinum data was collected using a TAP reactor to validate the methodology on real data. 
Details of the synthesis and the procedure of pretreatment of a 1.0 wt$\%$ $Pt/SiO_2$ catalyst can be found in our previous work \cite{constales2019methods}. 
Carbon monoxide oxidation on $Pt/SiO_2$ was carried out in a TAP-3 reactor. 
Approximately  15.6 mg of pretreated catalyst with the particle size of 250–300 $\mu m$ was loaded between two zones of the same particle size quartz sand (Sigma Aldrich). 
The TAP reactor was evacuated to a pressure of $10^{-7}$ torr and heated to $300^\circ C$ before the catalyst was subjected to at least three cycles of alternating pulses: 200 pulses of carbon monoxide / argon and 200 pulses of oxygen / helium to activate the platinum. 
Prior to the carbon monoxide oxidation experiment, the catalyst was again reduced at $300^\circ C$ by introducing a sequence of $50\%$ carbon monoxide / argon pulses until carbon dioxide formation was no longer detected. The TAP reactor was subsequently heated and kept at $450^\circ C$ for 30 minutes to remove adsorbed carbon monoxide and then cooled to the $150^\circ C$. 
The oxidation of carbon monoxide with oxygen on the catalyst was recorded by co-pulsing a 1:1 oxygen : helium mixture and a 1:1 carbon monoxide : argon mixture with the carbon monoxide delay of 0.01s. 
The time evolution of five mass fragments was followed, namely oxygen (atomic mass of 32), carbon dioxide (atomic mass of 44), carbon monoxide (atomic mass of 28), argon (atomic mass of 40), and helium (atomic mass of 4) with an Extrel quadrupole mass spectrometer, Extrel CMS, Pittsburg, PA. A total of 100 pulses per mass were injected into the reactor to investigate the kinetics.  

Each different flux response was baseline corrected and scaled by the calibration coefficient determined in an inert reactor.
This is done to ensure that the raw mass spectrometer measurements reflect the appropriate chemical signal.
Details of the TAP preprocessing workflow are described in Kunz et al \cite{kunz2018pulse}.
After the data is preprocessed, the gas concentration, reaction rate and uptake information are measured by the G-Procedure \cite{kunz2020probability}. 
A total of 50 pulses resulted in the detection of oxygen at the outlet flux where the RCDC matrix may be determined.

\section{Results and discussion}
To thoroughly examine the application of linear methods applied to a simple rate expression and a complete reaction mechanism, two different cases will be validated against simulated data. 
First, SCAD was  applied to simple elementary steps where the exact kinetic coefficients are known. 
This shows how SCAD was able to estimate the exact intrinsic kinetic coefficients while automatically selecting the correct gas/surface interactions. 
Secondly, both ER and LH mechanisms were simulated for the CO oxidation reaction. 
This will verify the ability to appropriately determine the mechanism in the presence of multiple interactions between active sites. 

\subsection{Rate expression validation}
A combination of the rate-reactivity model and penalized regression techniques are applied to each simulation in Section~\ref{sec:sim_data_elementary}.
Each of the simulated transient features are used to estimate the kinetic coefficients via the OLS, LASSO and SCAD  regression coefficients. 
Additionally, the ability to correctly choose the appropriate gas/surface interactions in the elementary step was measured by the negative prediction values (NPV) i.e. the ratio of the total number of true negatives and predicted negatives. 
A value of one for the NPV indicates a perfect selection of the coefficients, and hence the correct set gas/surface interactions within the given elementary step.   
The reported Root Mean Square Error (RMSE) is to be minimized for accurate prediction of the intrinsic kinetic coefficients. 
Table~\ref{tab:sim_results} displays the results of each method while reporting the NPV and the RMSE of the estimated intrinsic kinetic coefficient.

\begin{table}[h!]
	\centering
	\begin{tabular}{c || c c | c c | c c |} 
		\hline
		&\multicolumn{2}{|c|}{OLS}  &\multicolumn{2}{|c|}{LASSO} & \multicolumn{2}{|c|}{SCAD} \\
		\hline
		& NPV & RMSE & NPV & RMSE& NPV& RMSE\\
		\hline\hline
		1 & 1 &0.000 & 1 & 0.003& 1&0.000\\ 
		2a & 0 & 0.011 & 0.5 & 0.100&1&0.000 \\
		2b & 0 & 0.113 & 1 & 0.147&1&0.000 \\
		3 & 0 & 0.667 & 1 & 48.791&1&0.420 \\
		\hline
	\end{tabular}
	\caption{\label{tab:sim_results}The results of applying each regression method to each simulated reaction mechanisms in Table~\ref{tab:sim_coefficients}. }
\end{table}

The application of SCAD to the rate expression as a linear model results in the exact gas/surface interactions in each different basic elementary step denoted by the negative prediction value of 1. 
Additionally, SCAD was able to minimize the error in estimation of the kinetic coefficients.  
The  OLS model was effective in the prediction the kinetic coefficients but fails in distinguishing between relevant variables except for the most basic irreversible case. 
The LASSO model, while performing variable selection, only selects the proper variables in the simple irreversible simulation and fails to accurately estimate the kinetic coefficients when compared to OLS. 
The ability of SCAD to outperform both OLS and LASSO is due to the balance of variable selection while imposing less shrinkage on the regression coefficients  and hence better represent the kinetic rate constants. 

\subsection{Reaction mechanism kinetic coefficient estimation}

The irreversible ER mechanism for carbon monoxide oxidation can be described in order based on  the elementary steps 3 (second order irreversible) and 1 (irreversible)  in Table~\ref{tab:step_to_linear} as the following linear rate forms:

\begin{align}
	r_{O_2} &= \beta_{(N^2 k^+_O)} C_{O_2} - \beta_{(2Nk^+_O)} C_{O_2} U_O - \beta_{(k^+_O)} C_{O_2} U_O^2 \label{eq:ER_O_2} \\
	r_{CO} &= \beta_{(Nk^+_{CO})}C_{CO}U_O \label{eq:ER_CO}\\
	r_{CO_2} &= -r_{CO} \label{eq:ER_CO_2}
\end{align}
where the rate of the carbon dioxide product only depends on the rate of the carbon monoxide. 
Equations~\ref{eq:ER_O_2} and~\ref{eq:ER_CO} are not independent as they share the oxygen surface species. Replacing the oxygen surface species in Equation~\ref{eq:ER_CO}, the rate of carbon monoxide consumption can be written as:
\begin{equation}\label{eq:ER_RC}
	ER: \frac{r_{CO}}{C_{CO}} = \beta_{(Nk^+_{CO})} - \beta_{\left( k^+_{CO} / \sqrt{k^+_O} \right)}\sqrt{\frac{r_{O_2}}{C_{O_2}}}
\end{equation}
which can be solved as a single linear model for the entire mechanism. 
Note that the regression coefficients do not provide direct intrinsic kinetic estimates.  
Rather, the linear model only provides the apparent kinetic coefficient for carbon monoxide and the ratio of the intrinsic coefficients for the carbon monoxide and oxygen.

Another possible mechanism for the formation of carbon dioxide is through surface carbon monoxide interacting with the surface oxygen, i.e., the irreversible LH mechanism. 
This is based on the following gas rates of steps 3 (second order irreversible), 2 (first order reversible), and  4 (irreversible second order surface interaction) of Table~\ref{tab:step_to_linear} interacting with two different surface species. 
Each elementary step of the LH mechanism can be expressed in the linear form as:
\begin{align}
	r_{O_2} &= \beta_{(k^+_O)} C_{O_2} \left( N- U_O \right)^2 \label{eq:LH_O_2}\\
	r_{CO} &= \beta_{(k^+_{CO})}C_{CO} \left( N - U_O - U_{CO} \right) - \beta_{(k^-_{CO})} U_{CO} \label{eq:LH_CO} \\
	r_{CO_2} &= \beta_{(k^+_{CO_2})} U_O U_{CO} .
\end{align}
When each elementary step is assumed to be irreversible, i.e.,  the value of $\beta_{(k^-_{CO})}$ is equal to zero, then the kinetic coefficients for $k^+_{CO} / \sqrt{k^+_O}$ may be solved in the same manner as the ER mechanism in Equation~\ref{eq:ER_RC}, i.e.,
\begin{equation}\label{eq:LH_RC}
	LH: \frac{r_{CO}}{C_{CO}} = \beta_{\left( k^+_{CO} / \sqrt{k^+_O} \right)}\sqrt{\frac{r_{O_2}}{C_{O_2}}}
\end{equation}
However, due to the complex surface interactions with respect to the surface oxygen and carbon monoxide forming carbon dioxide, the intercept will no longer only be a scalar value of the number of active sites and the forward kinetic coefficient of carbon monoxide. 
As such, the intercept in Equation~\ref{eq:LH_RC} will provide a value of zero as both the rate of oxygen and carbon monoxide affect the total number of active sites. 
Additionally, there does not exist a simple linear relationship within the system of equations to determine the forward rate of carbon dioxide.

The SCAD method was applied to the rate-concentration dependency of the oxygen and carbon monoxide relationship detailed in Equations~\ref{eq:ER_RC} and~\ref{eq:LH_RC} directly from the simulated data  of each reaction mechanism. 
When each separate regression with a varying carbon monoxide kinetic coefficient was applied to the ER mechanism, the exact apparent rate coefficient for carbon monoxide $(Nk^+_{CO})$ was recovered from the intercept.
The slope $(k^+_{CO}/\sqrt{k^+_O})$ was estimated within a RMSE of $1.19 \times 10^{-14}$.
When the LH mechanism was applied with the same kinetic coefficients, the intercept was no longer significant in describing the rate-concentration dependency of carbon monoxide and was  set automatically to zero by the penalization.  
However, the slope $(k^+_{CO}/\sqrt{k^+_O})$ was estimated within an RMSE of $2.01 \times 10^{-15}$ showing the same ability to recover the kinetic coefficient as the ER mechanism. 
The difference in intercepts plays an important role in the linear correlation between all the rate/concentration ratios.

\subsection{Reaction mechanism correlation structure}

In the previous examples, we have demonstrated how the methodology can recover exact kinetic coefficients with a known simple mechanism. 
However, without  direct measurement of the conformation of surface species accumulation, the gas concentration and rates measured by the TAP reactor are insufficient to accurately describe the surface coverage. 
As such, the correlation between rate/concentration ratios found in Equations~\ref{eq:ER_RC} and~\ref{eq:LH_RC}  must be used in conjunction with the rate of carbon dioxide formation to determine the RCDC and mechanism structure. The ER mechanism initially relies on the accumulation of oxygen on the surface of the catalyst.
In this case, the rate/concentration dependence between the square root of oxygen and carbon monoxide must be correlated with a value of $-1$, i.e. carbon monoxide has a parasitic relationship with the surface oxygen. 
However, in the case of the LH mechanism, the rate/concentration dependencies between each reactant gas are mutualistic and will be positively correlated at a one-to-one relationship when assuming equal concentration. 
In the case where reversibility is significant in the adsorption of either gas species, explicit expressions such as Equations~\ref{eq:ER_RC} and~\ref{eq:LH_RC} cannot be derived but the correlations can be examined.  
In this case, the correlation between the oxygen and carbon monoxide rate-concentration dependency is diminished depending on the amount of reversibility. 

\begin{figure}[h!]
	\centering
	\includegraphics[width=.75\textwidth]{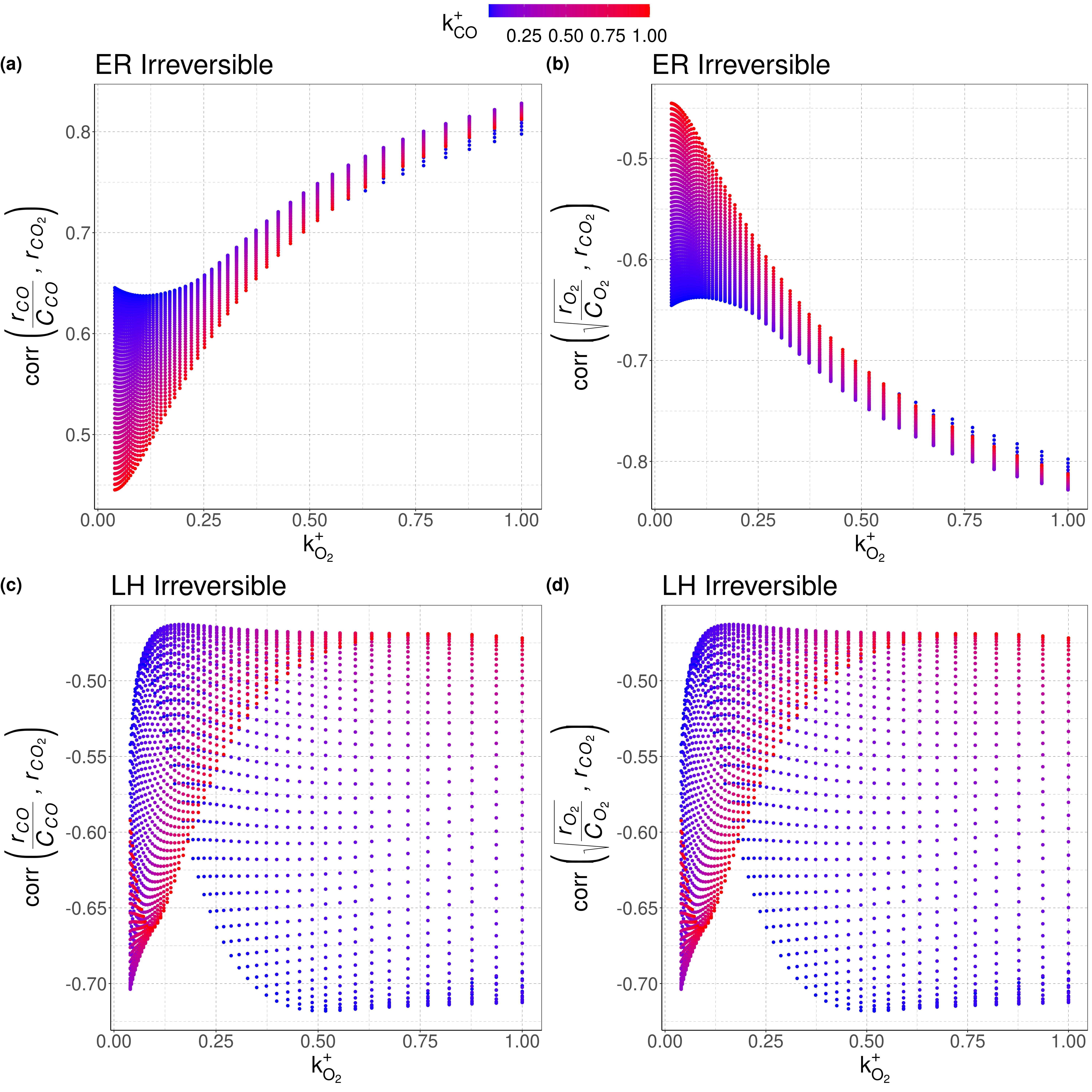}
	\caption{\label{fig:correlation_points_all}The correlation structure for Eley-Rideal and Langmuir-Hinshelwood when assuming an irreversible  process over a grid of potential forward rate coefficients. Sub figures (a) and (b) describe the rate/concentration dependency correlations of the Eley-Rideal mechanism of carbon monoxide and oxygen to the rate of carbon dioxide, respectively.  Similarly, (c) and (d) are the rate/concentration dependency correlations with respect to the Langmuir-Hinshelwood mechanism for each reactant. } 
\end{figure}

For the irreversible case, the correlations were  determined numerically by looking at the RCDC over a series of simulated kinetic rate coefficients found in Section~\ref{sec:sim_data_mechanism}.  
Figure~\ref{fig:correlation_points_all} compares the results of RCDC for each mechanism.
The correlation coefficients of oxygen,$\sqrt{r_{O_2}/C_{O_2}}$, and carbon monoxide,$r_{CO}/C_{CO}$, were compared to carbon dioxide, $r_{CO_2}$, over a set of simulated forward rate coefficients , i.e. $k_CO^+$ and $k_{O_2}^+$ range in value from 0.04 to 1.00 $m^2/mol\/s$. 
The ER mechanism in (a) and (b) of Figure~\ref{fig:correlation_points_all} exhibits a non-linear behavior with respect to the increase in the forward rate constant of oxygen. 
This is due to the assumed second order effect in the elementary step of oxygen accumulation. 
This also affects the spread of the correlations with an increase in the forward rate constant of carbon monoxide due to the dependency of the oxygen surface concentration and carbon dioxide. 
When both reactants in (a) and (b) are compared, the only difference exists is the sign of the correlation coefficient. 
This sign change is not apparent in the LH mechanism due to the mutualistic behavior of the oxygen and carbon monoxide surface species, seen in (c) and (d) of Figure~\ref{fig:correlation_points_all}. 
As the oxygen reactant rate constant is maximized and the carbon dioxide rate constant is minimized (or vice versa), the correlation to the rate of carbon dioxide approaches -1. 
Additionally, the non-linear behavior of the correlation coefficient is due to the combination of the first order carbon monoxide and second order oxygen surface effects. 
Note that the correlation structures in the Eley-Rideal mechanism between each reactant differ only by the sign over the grid of forward kinetic coefficients.

\begin{figure}[h!]
	\centering
	\includegraphics[width=.8\textwidth]{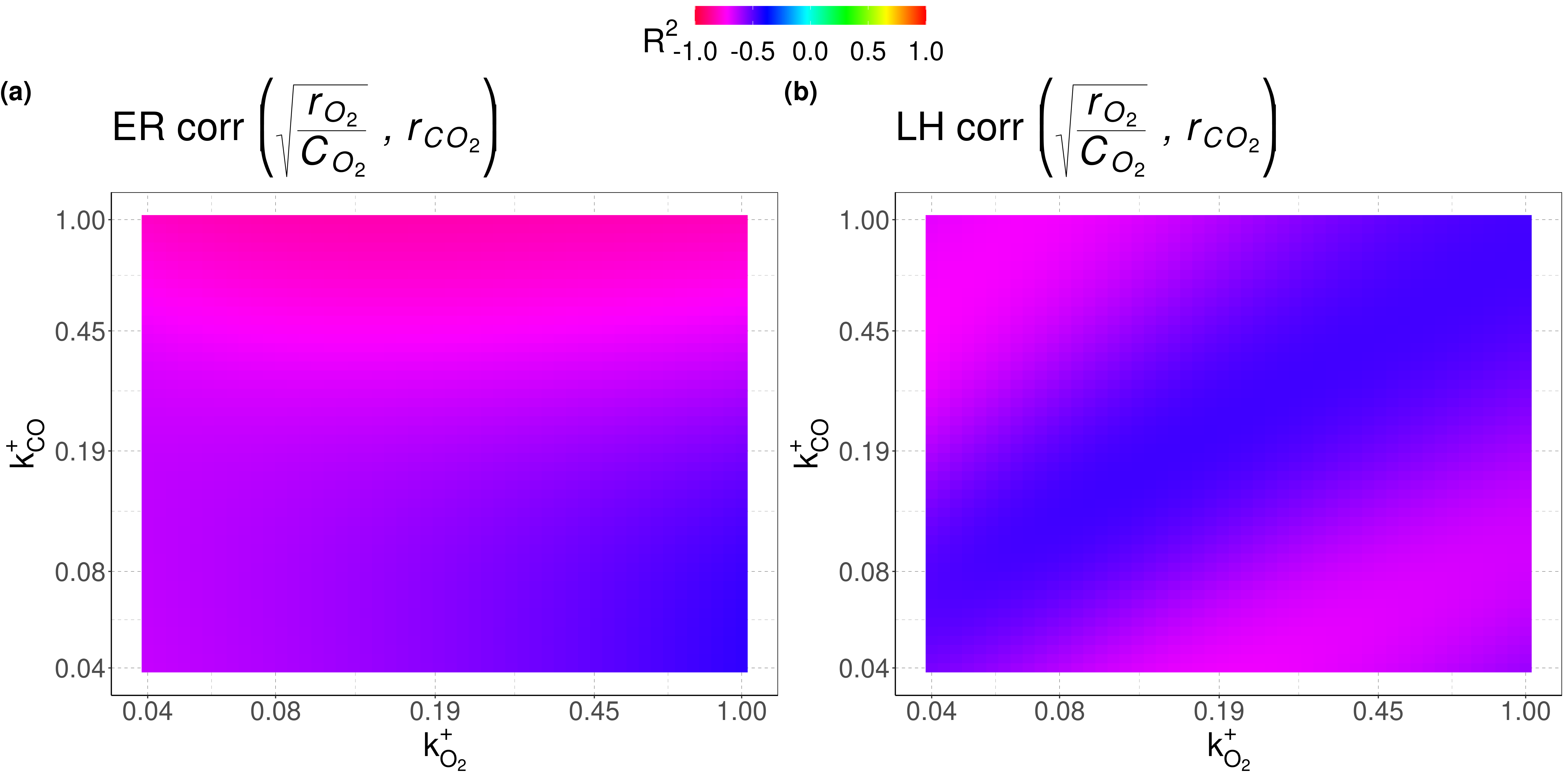}
	\caption{\label{fig:correlation_heat_irreversible}The comparison of the correlation coefficients for oxygen ($\sqrt{r_{O_2}/C_{O_2}}$) to $CO_2$ over the Eley-Rideal (a) and Langmuir-Hinshelwood (b) mechanisms.} 
\end{figure}

Further distinction between the ER and LH mechanisms can be made by examining the correlation between the reactants and products as a heat map in Figure~\ref{fig:correlation_heat_irreversible}. 
For the ER mechanism, as the carbon monoxide forward rate constant increases, the rate/concentration dependency of both the reactants are more correlated with the rate of carbon dioxide. 
This aligns with the linear regression analysis where the rate of carbon dioxide is directly impacted by the rate of carbon monoxide and the number of active sites. 
On the other hand, the LH mechanism which relies on the surface species interactions, produces a correlation coefficient of -0.5 along the diagonal of the oxygen and carbon monoxide forward rate coefficients. 
Moving away from the diagonal demonstrates a non-linear behavior increasing the correlation coefficient to the products as a single reactant becomes more dominant in the mechanism . 
Note how the reactants in the ER mechanism are highly correlated with the carbon dioxide as the forward rate of carbon monoxide increases signifying dependence on the surface oxygen. 
Additionally, the reactants in the Langmuir-Hinshelwood exhibit a correlation coefficient of -0.5 when the forward rate of oxygen and carbon monoxide are equal. 
This is due to the nature of the mechanism where the active sites are competing to fill the surface of the catalyst. 

\begin{figure}[h!]
	\centering
	\includegraphics[width=.8\textwidth]{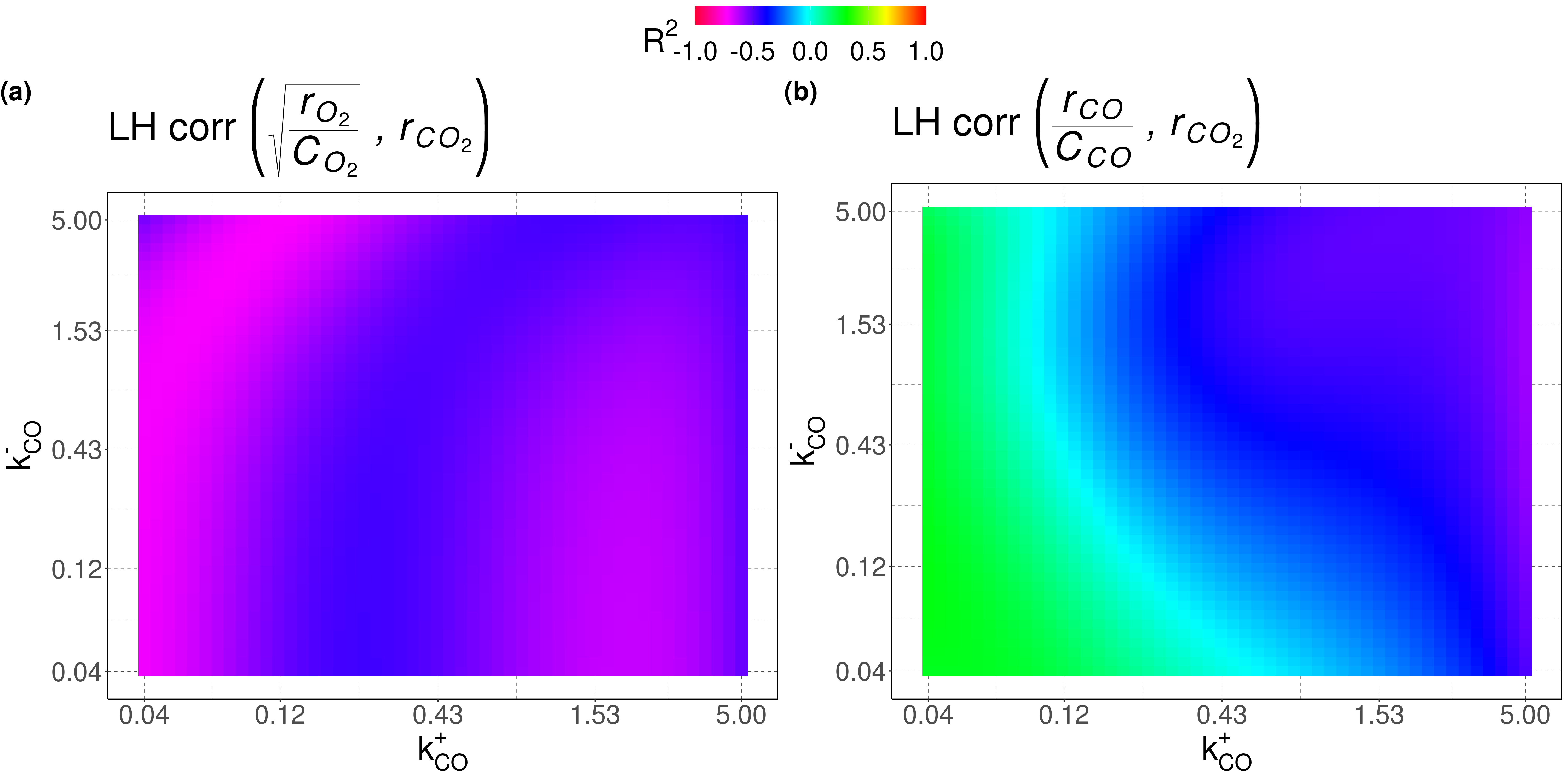}
	\caption{\label{fig:correlation_heat_reversible}The comparison of the absolute value correlation coefficients for the carbon dioxide rate compared to the oxygen ($\sqrt{r_{O_2}/C_{O_2}}$) (a) and carbon monoxide ($r_{CO} / C_{CO}$) (b) when reversibility is present in the Langmuir-Hinshelwood mechanism. } 
\end{figure}

Next, examining the effect of reversibility on carbon monoxide, Figure~\ref{fig:correlation_heat_reversible} describes the correlation structure of the oxygen (a) and carbon monoxide (b) compared to carbon dioxide over the range of forward and reverse rate constants for CO in the LH mechanism. 
Both sub-figures plot the relationship of the square root of the rate-concentration dependency as oxygen has a second order interaction with the surface of the catalyst, see Equation~\ref{eq:ER_RC} and Equation~\ref{eq:LH_RC}.
The oxygen RCDC heatmap shown in part (a) describes two potential areas where there exists a large negative correlation with carbon dioxide.  
This corresponds to bounds on which the forward rate constant may exist, approximately 0.04 to 0.08 and 0.43 to 5  $m^2/mol\/s$.  
This is perhaps more informative when collaborated with the RCDC of carbon monoxide to carbon dioxide in Figure~\ref{fig:correlation_heat_reversible} (b).  
Correlation values of the carbon monoxide RCDC of 0.5 indicate that the forward rate constant of carbon monoxide ranges from approximately 0.04 to 0.43 $m^2/mol\/s$. 
As the forward rate constant of carbon monoxide approaches the static forward rate constant of carbon dioxide of 5.0 $m^2/mol\/s$, the correlation value decreases toward negative one.   
As such, a reversible LH mechanism can be discerned from the irreversible case by the difference in the correlation of oxygen and carbon monoxide with respect to carbon dioxide. 

A detailed analysis of this simulation data provides the means of determining the specific set of elementary steps that may be apparent in experimental correlation structures. 
The differences in the ER and LH mechanism are due to the gas phase carbon monoxide interacting with the surface oxygen or both species interacting on the surface of the catalyst, respectively.
As such, the ER mechanism carbon monoxide will have a positive correlation to carbon dioxide and the LH mechanism carbon monoxide will have a negative correlation to carbon dioxide.
If the correlation coefficients of the reactants agree in magnitude and disagree in sign, then the ER mechanism is apparent. 
Conversely, complete reactant correlation coefficient agreement would suggest an irreversible LH mechanism. 
This can be further investigated by comparing the reactants correlation structure to the reversible LH mechanism. 
Finally, while the coefficient structure does not give traditional micro-kinetic coefficient estimates, it can be used to determine bounds for a given catalyst. 

\subsection{Reaction mechanism estimation from experimental data}
\begin{figure}[h!]
	\centering
	\includegraphics[width=1\textwidth]{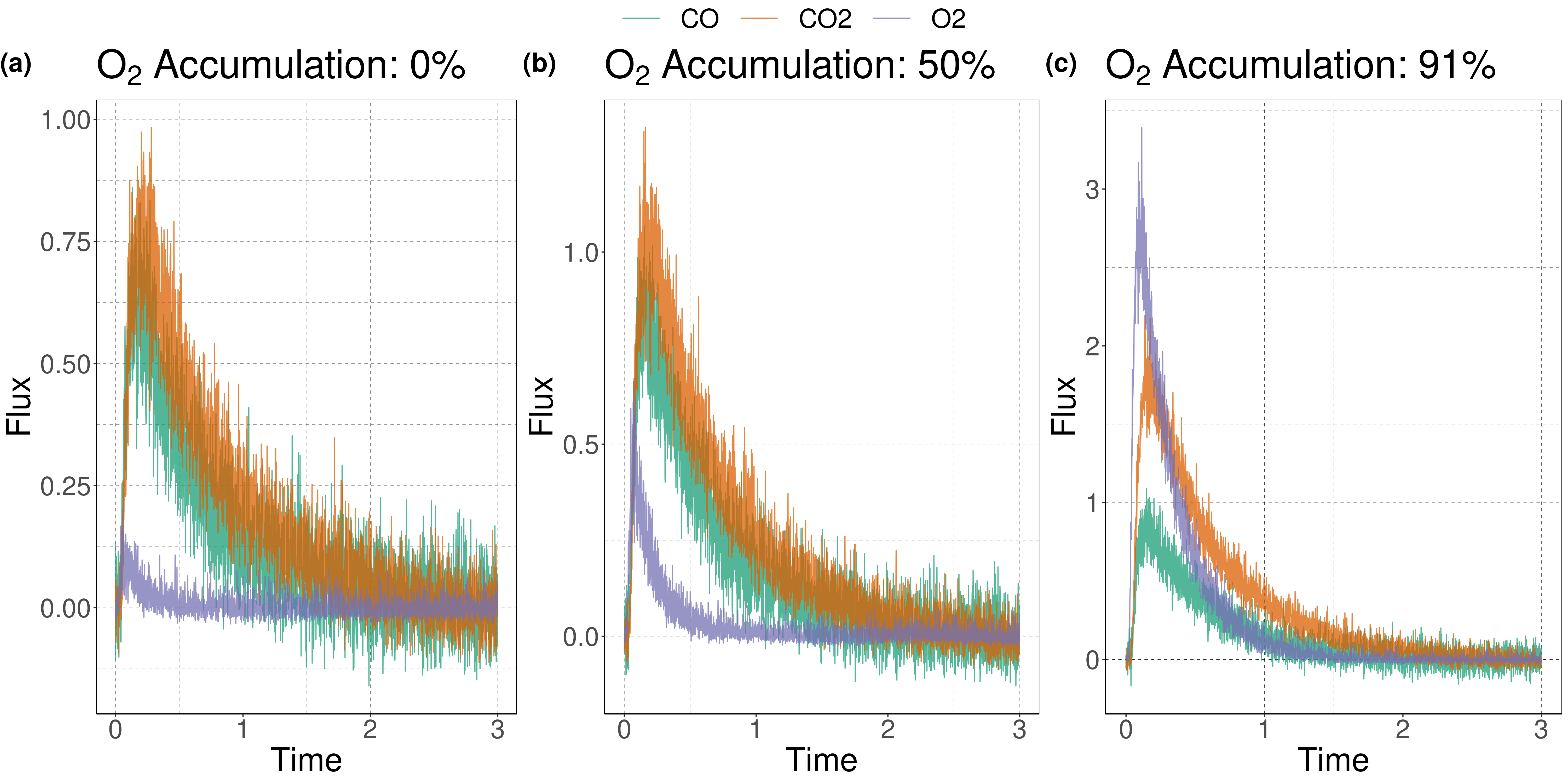}
	\caption{\label{fig:pt_experimental_pulse_series}The outlet flux responses measured by the TAP reactor over a series pulses.  Each sub figure displays a different state of oxygen accumulation at the beginning, middle and end of the TAP experiment.} 
\end{figure}

Application of the RCDC structure to experimental data requires special attention to the amount of experimental noise in the system. 
As demonstrated with the CO oxidation example, all gas species responses must be measurable compared to the noise to obtain valid insight into the reaction mechanism. 
The experimental carbon monoxide oxidation experiment consists of 50 pulses where flux was measured and rate/concentration dependencies for all gases were calculated.  
Each pulse may be considered a separate experiment within the same environmental conditions where pressure and temperature are constant.  
Starting from a completely reduced catalyst, as the surface accumulation of oxygen changes over each pulse, measured as the surface oxygen uptake $(\int_0^t 2r_{O_2} - r_{CO_2}dt)$. 
Figure~\ref{fig:pt_experimental_pulse_series} displays the experimental outlet flux responses for each gas at different oxygen accumulation states. As the total number of actives sites are filled with oxygen, the oxygen and carbon dioxide outlet flux become larger in magnitude. 
To clearly determine a change in the mechanism, the outlet responses must be transformed to their respective transient kinetic features via the Y or G-Procedure. 
Furthermore, the noise in the system may cause significant outliers when determining the rate-concentration dependencies and hence cause poor estimation of the correlation structure.  
As such, robust covariance estimation must be leveraged to obtain accurate estimation of the RCDC. 

\begin{figure}[h!]
	\centering
	\includegraphics[width=.5\textwidth]{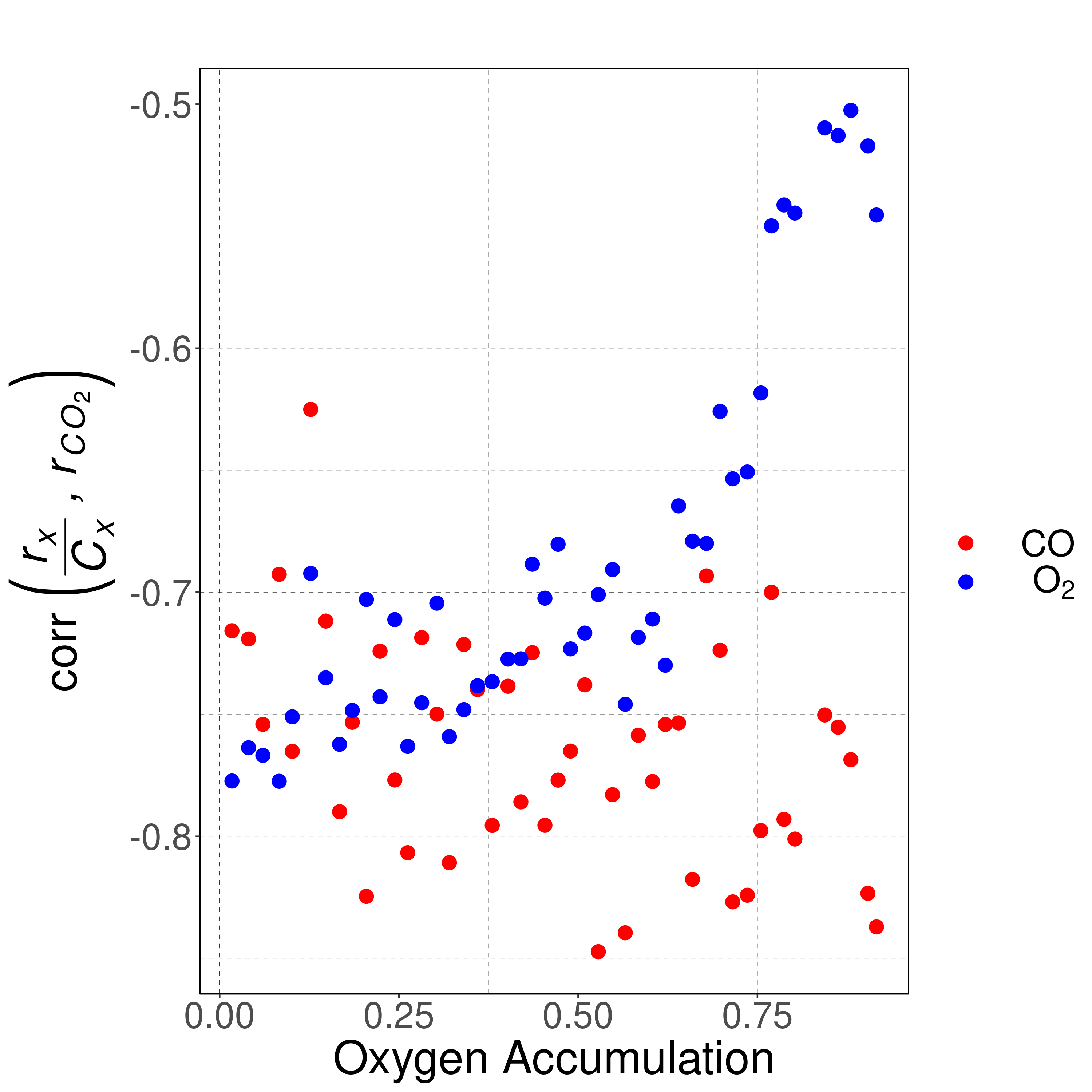}
	\caption{\label{fig:platinum_correlation_over_pulse}The RCDC of carbon monoxide and oxygen to carbon dioxide, as $x$ depending on coloration, over the change of oxygen accumulation in experimental data. The correlation agreement in the first 25$\%$ of the oxygen accumulation describes a Langmuir-Hinshelwood mechanism dominated by either the forward kinetics of oxygen or carbon monoxide. As the oxygen accumulation increases, the correlation drifts describe a larger impact on the forward rate of carbon monoxide to the mechanism.} 
\end{figure}

The RCDC of oxygen and carbon monoxide to carbon dioxide is plotted in Figure~\ref{fig:platinum_correlation_over_pulse}. 
With both correlation coefficients being negative, this is clear sign from Figure~\ref{fig:correlation_points_all} that the experimental data is best described by an LH mechanism. 
The correlation coefficients of both species are roughly equal and do not change within the first 25$\%$ of oxygen accumulation.  
Based on the correlation of oxygen, the forward and reverse rate constants of carbon monoxide in low oxygen surface accumulation exhibit similar behavior to the “island” of correlation values in Figure~\ref{fig:correlation_heat_reversible} (a).
As such, the forward rate constant of carbon monoxide could be less than 0.04 $m^2/mol\/s$ or range from 0.43 to 5 $m^2/mol\/s$.
Since the oxygen correlation values approach -0.5, the forward rate of carbon monoxide can either be transitioning off of these islands within ranges of 0.12 to 0.43 $m^2/mol\/s$ or be greater than 5 $m^2/mol\/s$. 
The carbon monoxide correlation to carbon dioxide decreases approaching -1 as oxygen accumulates on the surface.
Inspection of Figure~\ref{fig:correlation_heat_reversible} (b) reveals that this large negative correlation is well described only when the forward rate of carbon monoxide approaches 5 $m^2/mol\/s$ while the reverse rate of carbon monoxide is unknown.

The correlation values obtained from the RCDC information were collected per pulse with respect to a specific catalyst surface state. 
As such, each pulse gives insight to the mechanism individually and can be thought of as an individual experiment. 
As each pulse further perturbs the surface of the catalyst, the correlation values indicate a shift in the apparent kinetics. 
With a negative correlation of carbon monoxide, the ER mechanism may be excluded as the only mechanism. 
However, both mechanisms may still have an impact on the correlation coefficients as the oxygen accumulates.  
This transition to a different mechanism could be justified through examples in literature where it is proposed that two reaction models may exist with different adsorption, desorption and surface reactions based on environmental and surface conditions \cite{medford2019scalable, liu2017dense, herz1980surface}. 
This highlights the ability of the TAP technique to collect a vast amount of kinetic information as the interplay between the gas and surface concentration incrementally evolves. 
A similar design of  experiments performed using steady-state techniques would be extremely laborious. 

\section{Conclusion}

The rate-reactivity model is a distinct method for bridging the traditional micro-kinetic model approach to experimental transient analysis on complex, multi component catalysts that are not readily amenable to atomistic simulations.  
However, the interpretation of the reactivity coefficients requires examples from clearly defined simulation.
This paper addresses some of the unique features of the rate-reactivity model by directly showing that if an experiment involves a single elementary step and single type of active site, then the intrinsic kinetic coefficients may accurately be obtained with the use of non-convex variable selection in machine learning. 

When the reaction mechanism consists of multiple elementary steps, the correlation structure of the reactants and products may be used to classify a given mechanism.  
This was validated against common simple mechanisms, i.e. the Eley-Rideal and Langmuir-Hinshelwood, and applied to the carbon monoxide oxidation using TAP pulse response data simulated from each micro-kinetic model. 
Examining multiple combinations of kinetic coefficients via simulated data yields information that may be used in determining bounds on the kinetics coefficients for experimental data. 
This information can then be used as initial values in detailed simulation efforts to understand experimental data \cite{yonge2020tapsolver}.  
Further extensions will be focused on exploring more complex mechanisms where interactions between certain gas species may not exist.  
In this case, penalized graph structures will be used in determining the significant gas interactions in a data driven manner without any $a priori$ assumption of the reaction mechanism.

The given methodology was applied to experimental data for CO oxidation on platinum where the Langmuir-Hinshelwood mechanism was readily apparent. 
The experimental data demonstrates that the RCDC was able to invalidate the hypothesis that the Eley-Rideal mechanism is the only or the dominating mechanism.
Additionally, combining experimental and simulation results, bounds for the rate constants were obtained. 
As oxygen accumulation increased on the surface of the catalyst, the apparent kinetics were altered based on the availability for the carbon monoxide gas concentration interaction with the catalyst surface. 
As a data driven method, the RCDC is able to minimize chemical bias and intuition in characterizing the performance of a given industrial catalyst. 
As such, it is the goal to generalize the techniques developed in this paper to apply to any reaction for greater understanding about complex catalysts, a reduction in computation time, and determining the root causes for the behavior of a given catalyst. 

\section{Acknowledgments}
Support for this work was provided by the U.S. Department of Energy (USDOE), Office of Energy Efficiency and Renewable Energy (EERE), Advanced Manufacturing Office Next Generation R$\&$D Projects under contract no. DE-AC07-05ID14517. We thank Dr. Yixiao Wang, Dr. Rakesh Batchu and Mr. James Pittman for many supportive discussions.

\bibliographystyle{ieeetr}
\bibliography{tap.bib}

\end{document}